\documentclass[article,nojss]{jss}
\usepackage{framed,longtable,amsmath,amsfonts,amssymb,orcidlink,thumbpdf,lmodern}
\shortcites{chenxgboost,datta2019,hersbach2023pl,hersbach2023sfc}

\title{Long-term foehn reconstruction combining unsupervised and supervised learning}

\author{Reto Stauffer~\orcidlink{0000-0002-3798-5507}\\Universit\"at Innsbruck
   \And Achim Zeileis~\orcidlink{0000-0003-0918-3766}\\Universit\"at Innsbruck
   \And Georg J. Mayr~\orcidlink{0000-0001-6661-9453}\\Universit\"at Innsbruck}
\Plainauthor{Reto Stauffer, Achim Zeileis, Georg J. Mayr}

\Abstract{
  Foehn winds, characterized by abrupt temperature increases and wind speed
  changes, significantly impact regions on the leeward side of mountain
  ranges, e.g., by spreading wildfires. Understanding how foehn occurrences
  change under climate change is crucial. Unfortunately, foehn cannot be
  measured directly but has to be inferred from meteorological measurements
  employing suitable classification schemes. Hence, this approach is typically
  limited to specific periods for which the necessary data are available.

  We present a novel approach for reconstructing historical foehn occurrences
  using a combination of unsupervised and supervised probabilistic statistical
  learning methods. We utilize in-situ measurements (available for recent
  decades) to train an unsupervised learner (finite mixture model) for
  automatic foehn classification. These labeled data are then linked to
  reanalysis data (covering longer periods) using a supervised learner (lasso
  or boosting). This allows to reconstruct past foehn probabilities based
  solely on reanalysis data.

  Applying this method to ERA5 reanalysis data for six stations across
  Switzerland and Austria achieves accurate hourly reconstructions of north
  and south foehn occurrence, respectively, dating back to 1940. This paves the way
  for investigating how seasonal foehn patterns have evolved over the past
  83~years, providing valuable insights into climate change impacts on these
  critical wind events.
}

\Address{
  Reto Stauffer\\
  Universit\"at Innsbruck\\
  Faculty of Economics and Statistics\\
  \emph{and} Digital Science Center\\
  Universit\"atsstr.~15\\
  6020 Innsbruck, Austria\\
  E-mail: \email{Reto.Stauffer@uibk.ac.at}
}

\begin{document}

% ===================================================================
% ===================================================================
% ===================================================================
\section{Introduction}

Foehn winds are downslope winds on the leeward side of mountains and can be
found all around the world in areas with pronounced topographical features that
impede the airflow, such as the European Alps, the Southern Alps in New
Zealand, mountain ranges along the Mediterranean Sea, or the Rocky Mountains.
Depending on the region, these winds are given specific names
such as Santa Anna winds (Southern California), Chinook (Rocky Mountains),
Mistral (Southern France), Bora (Croatia), or Foehn (Central Europe and
New Zealand). For historical reasons `foehn' has become a synonym for
this type of terrain-induced wind phenomena.

Often, foehn is characterized by a sharp increase in wind speed and sudden
changes in temperature and relative humidity, which can have a strong influence
on the local climate and the people living in the affected areas.
While foehn is often associated with a mild (and typically dry)
climate, strong foehn events can also cause extensive damage to vegetation
and man-made structures. In some areas it is not uncommon for strong foehn gusts to
overturn trucks or vans, or for airports and harbors to be closed due to
unsafe conditions. In addition, the strong and dry winds can kindle
and spread domestic fires and wild fires
\citep{schoennagel2004,reinhard2005,zumbrunnen2009} or affect the development
of Antarctic ice shelves (e.g., \citealt{cape2015,elvidge2020}).

Although the conceptual model of foehn is well established
\citep{armi2007,mayr2008,armi2011,richner2013}, it cannot be measured directly.
During the last decades several (semi)-automatic methods have been developed
which allow to differentiate `foehn' and `no-foehn' events based on in-situ
measurements from automated weather stations (AWSs). Among the frequently used
algorithms are Widmer's f\"ohn index \citep{widmer1966,courvoisier1971}
based on Fisher's linear discriminant analysis to distinguish between two or
more distinct classes, and enhanced versions of it (e.g.,
\citealt{jansing2022}). Other studies use decision-based or tree-based methods
to classify foehn events
(e.g., \citealt{duerr2008,spiers2013,cape2015,turton2018,datta2019,elvidge2020,laffin2021,francis2023,laffin2021,francis2023}),
all of which are deterministic methods, where the thresholds have often
been selected manually. To overcome these limitations,
\cite{plavcan2014} proposed a method based on finite mixture models for automatic
and fully parametric probabilistic foehn classification. To perform the classification, all methods
require AWS measurements with high temporal resolution (ideally sub-hourly),
which are typically only available for recent
decades. While this allows to classify and analyze foehn when the AWS
provides sufficient data, there is no information on foehn prior to the
installation of the AWS, nor when there have been outages or the AWS has been
decommissioned.

Additional information on the atmospheric conditions from numerical
(re-)analysis models are an excellent source to complement the in-situ
measurements. Reanalysis data sets are typically produced by physically
based numerical weather prediction (NWP) models and sophisticated data assimilation
schemes that use all available observations to estimate the
`best known' atmospheric state. An example is the global reanalysis
data set ERA5 \citep{hersbach2023pl,hersbach2023sfc} from the European Centre
for Medium-Range Weather Forecasts (ECMWF). ERA5 provides
global hourly 4-dimensional atmospheric conditions back to 1940.
However, this comes with its own challenges. First, foehn is not explicitly
modeled by the NWP models. Second, due to technical and computational
limitations, the reanalysis can only approximate the real world and cannot
resolve small-scale atmospheric processes and topographic features, which
would be important for small-scale phenomena such as foehn.

One way to overcome these limitations is to combine AWS measurements and reanalysis
data using statistical or machine learning techniques. A
classification of foehn based on AWS data serves as the response
(target/outcome/labels) for a supervised model that uses reanalysis data
as explanatory variables (inputs/covariates). Once the relationship
between the two sets of data has been learned, the models can be used to predict
the expected state (`foehn' or `no foehn') for periods for which reanalysis data
is available but AWS measurements are not. This is also known as (statistical)
downscaling or post-processing and has been used for foehn modeling in some
variations: For nowcasting foehn at a station in Switzerland (Altdorf),
\cite{sprenger2017} use data from a local 7\,km analysis data set (COSMO-7) --
but only to get consistent inputs for their adaptive boosting algorithm (AdaBoost).
\cite{laffin2021} use both ERA5 and the Regional
Atmospheric Climate Model 2 (RACMO2; \citealt{racmo2}) combined with tree-based
gradient boosting (XGBoost; \citealt{chen2016}) to predict foehn on the
Antarctic Peninsula. XGBoost is also used by \cite{mony2021} to investigate
future changes to foehn frequency in Switzerland.

This study proposes a novel probabilistic approach that combines unsupervised
and supervised machine learning methods to bridge the gap between in-situ
Automatic Weather Station (AWS) measurements and ERA5 reanalysis data to diagnose
foehn. This combined approach allows us to reconstruct long-term, high-resolution
foehn conditions over several decades. ERA5 data enables us to reconstruct the
probability of foehn occurrence with hourly temporal resolution dating back to
1940, long before AWS were installed.

Our approach also allows to identify potential long-term changes in foehn
occurrence in the European Alps from this high-resolution reconstruction.
The validity of this combined approach is demonstrated by applying it to six
stations located both north and south of the main Alpine ridge to show the
method's effectiveness for both north and south foehn wind situations.

\begin{figure}[!ht]\centering
    \includegraphics[width=\textwidth]{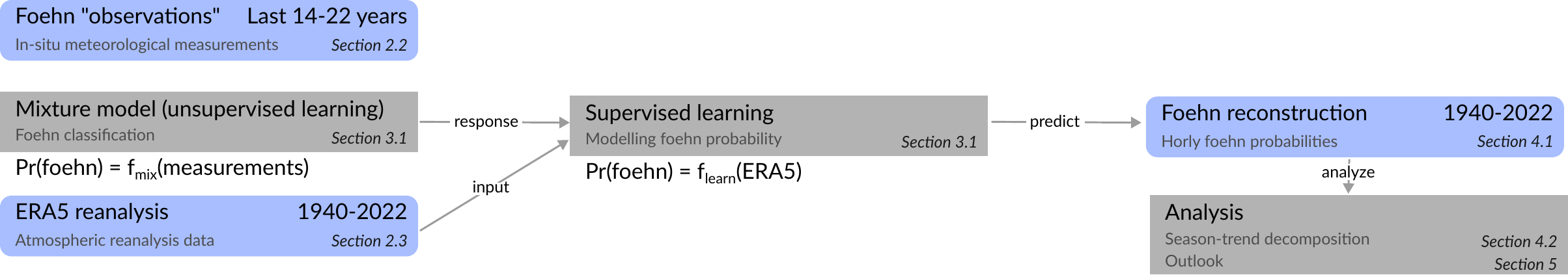}
    \caption{\label{fig:schemata} Flow chart for the new combined foehn
    reconstruction algorithm.}
\end{figure}

Figure~\ref{fig:schemata} shows a schematic representation of the proposed approach.
For all six locations, data are available from an AWS at the target
location as well as from a nearby mountain station for the last 14--22 years
(Sec.\,\ref{sec:data:obs}) on a 10\,min temporal interval. A Gaussian mixture
model (unsupervised learning; Sec.\,\ref{sec:method:mixture}) is used for foehn
classification. The result is then aggregated into binary time series
(`foehn'/`no foehn') with an hourly temporal resolution to match the resolution
of the ERA5 data used in the next step. After combining the different data sources, supervised
learning (Sec.\,\ref{sec:method:glm}) is used to find the relationship between
a variety of interpolated and derived variables from ERA5
(Sec.\,\ref{sec:data:era5}) and the classified events. Once these statistical
models have been estimated, foehn can be reconstructed
(Sec.\,\ref{sec:method:reconstruction}) for the whole period from 1940--2022,
allowing to investigate possible trends and/or seasonal changes over the last
decades (Sec.\,\ref{sec:method:str}).

% ===================================================================
% ===================================================================
% ===================================================================
\section{Data}\label{sec:data}

Section\,\ref{sec:data:obs} describes the measurement data utilized for foehn
classification along with the study area and the target stations.
Section\,\ref{sec:data:era5} explains the reanalysis data set and its
pre-processing for the supervised learning method, along with the
reconstruction process.

% -------------------------------------------------------------------
\subsection{In-situ measurements}\label{sec:data:obs}

This study utilizes data from six AWSs
situated across Switzerland and the western part of Austria, all positioned at the
bottom of valleys known to be affected by foehn winds. Four of these
stations are located north of the main Alpine ridge, while two are
located in the canton of Ticino (Switzerland) south of the main
Alpine ridge. Whilst the stations north
of the Alps are prone to south foehn, the stations south of the Alps are known
for the presence of north foehn.

An additional AWS upstream near the crest of the main Alpine range
improves the accuracy of the foehn classification \citep{plavcan2014}.
The stations Innsbruck and Ellb\"ogen utilize data from
Sattelberg, the remaining four stations in Switzerland use observations from
station G\"utsch.

All stations provide data on mean wind speed, wind direction,
air temperature, and relative humidity at a
10\,min temporal resolution. Table\,\ref{tab:stations} displays the locations
and data availability of these stations, while Figure\,\ref{fig:neighbors}
depicts a map illustrating their geographical position and the surrounding
topography.

% latex table generated in R 4.3.1 by xtable 1.8-4 package
% Mon Jun  3 12:25:05 2024
\begin{table}[ht]
\centering
\begin{tabular}{llll}
  \hline
 & Type & Location & Data availability \\ 
  \hline
$\bigtriangleup$  Gütsch (Andermatt)$^1$ & crest & 46.653N/8.616E 2286m & 2005-01-01--2023-12-31 (95.3\%) \\ 
  Altdorf$^1$ & south & 46.890N/8.620E 438m & 2005-01-01--2022-12-31 (78.1\%) \\ 
  Montana$^1$ & south & 46.290N/7.460E 1423m & 2005-01-01--2022-12-31 (77.1\%) \\ 
  Comprovasco$^1$ & north & 46.460N/8.935E 576m & 2005-01-01--2022-12-31 (85.8\%) \\ 
  Lugano$^1$ & north & 46.004N/8.960E 205m & 2005-01-01--2022-12-31 (90.2\%) \\ 
  $\bigtriangleup$  Sattelberg$^2$ & crest & 47.011N/11.479E 2107m & 2006-01-01--2022-12-31 (75.0\%) \\ 
  Ellbögen$^2$ & south & 47.200N/11.430E 1080m & 2006-01-01--2022-12-31 (92.0\%) \\ 
  (Universit{\"a}t) Innsbruck$^3$ & south & 47.260N/11.385E 578m & 2009-06-21--2022-12-30 (99.1\%) \\ 
   \hline
\end{tabular}
\caption{Station type, location and data availability; begin and end date plus
percent available within period.
Observations provided by the Swiss national weather service (1; MeteoSwiss), the University of
Innsbruck (2) and the Austrian national weather service (3; GeoSphere Austria).
Four stations are used to model
south foehn, two to model north foehn and two
serving information at the mountain crest ($\bigtriangleup$; cf. Type).
} 
\label{tab:stations}
\end{table}

% -------------------------------------------------------------------
\subsection{ERA5 reanalysis}\label{sec:data:era5}

%To be able to reconstruct foehn occurrences for period lacking in-situ
%measurements, the atmospheric conditions at that time have to be known.
This study makes use of ERA5 reanalysis data, which is publicly accessible via the
Copernicus climate data store \citep{hersbach2023pl,hersbach2023sfc}.
ERA5 offers four-dimensional gridded data with an hourly temporal resolution
(starting from 1940) on a spatial grid of $0.25^\circ \times 0.25^\circ$ ($\sim 28\,km
\times 20\,km$ for Central Europe).

%% Preliminary? Quick check shows that 1978-xx has been downloaded in March,
%% Thus already the updated data set? Hard to tell ...
%% https://www.ecmwf.int/en/newsletter/175/news/era5-reanalysis-now-available-1940
%% YES DOWNLOADED AFTERWARDS.

\begin{figure}[!ht]\centering
    \includegraphics[width=0.7 \textwidth]{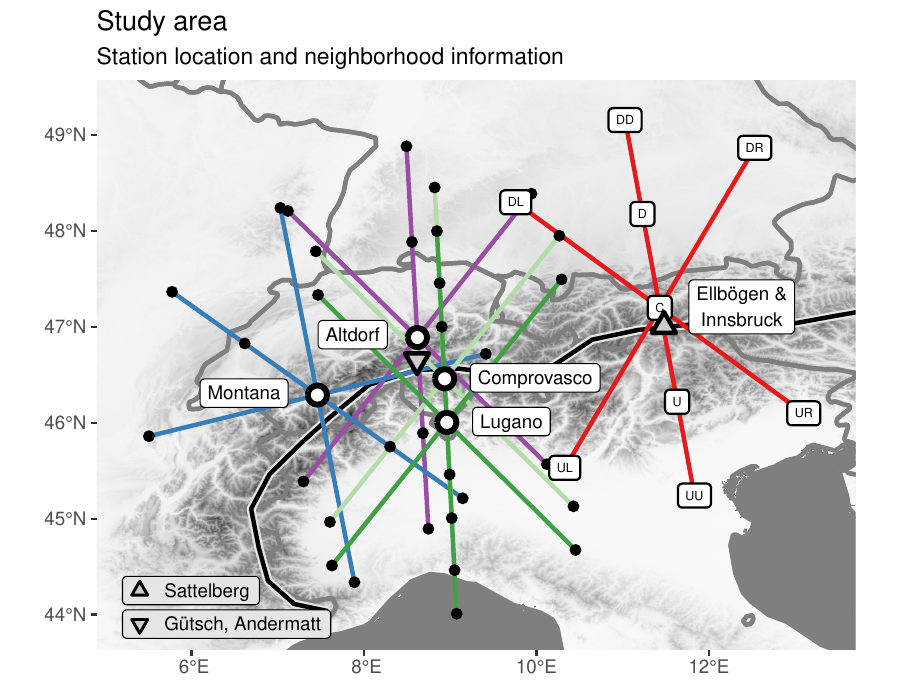}
    \caption{\label{fig:neighbors}
        Location of the six target stations (circles/C) and two crest stations
        (triangles); Innsbruck and Ellb\"ogen are shown combined due to their
        close proximity ($<8\,km$).
        The solid black lines represents the main Alpine ridge.
        On top, the neighborhood `star' used for interpolation is shown, exemplarily labeled
        on the most eastern station (details see App.\,\ref{app:sec:data:era5}).
    }
\end{figure}

The data from 90 different fields (30 single level fields and 60 pressure level
fields; see App.\,\ref{app:sec:data:era5}) are bilinearly interpolated to the
geographical location of the six stations of interest. Based on the 90 
interpolated values, a series of derived variables is calculated, including
vertical temperature gradients and level thickness, resulting in a total of 155
variables. These 155 variables are referred to as the `direct' variable set as they
solely rely on information retrieved directly from the geographical location of the
corresponding target station.

Since foehn is driven by large-scale synoptic processes, it might be insufficient
to rely only on data at the target location. Therefore,
additional information from the surrounding is incorporated by extracting data
from ERA5 at a series of neighboring points in a `star' formation around the
target location. In combination with the direct variables, a list of additional
derived variables is calculated, such as spatial temperature differences and
spatio-temporal pressure differences, among many others. Finally, the first
and second order harmonics of the day of the year are included to capture
seasonal variation. In total, this yields 497 variables:
four harmonics, 155 direct variables, 136 spatial variables, 120
temporal variables and 82 spatio-temporal variables (e.g., temporal
changes of spatial differences). This expanded set is referred to as the `full'
variable set.

Figure\,\ref{fig:neighbors} shows the target locations (\verb|C|; center) and
their neighboring points used. These neighboring points are positioned
geographically relative to the target station upstream (\verb|U|) and
downstream (\verb|D|) of the main foehn wind direction as well as to the right
(\verb|R|) and left (\verb|L|) of it. While the interpolated information from
the target location itself (\verb|C|) is always used as possible covariate for
the statistical models, the values interpolated at the neighboring points are
not directly employed but are instead used for the calculation of the
derived/augmented variables. Further details regarding the
construction of the neighboring `star' can be found in
Appendix\,\ref{app:sec:data:era5}.

% ===================================================================
% ===================================================================
% ===================================================================
\section{Methodology} \label{sec:method}

Section\,\ref{sec:method:mixture} introduces the unsupervised learning model
used for foehn classification, followed by the supervised learning models in
Section\,\ref{sec:method:glm}. The results from Section\,\ref{sec:method:glm}
are then used to reconstruct hourly foehn occurrence over the past decades
(Sec.\,\ref{sec:method:reconstruction}) which is analyzed employing season-trend
decomposition in Section\,\ref{sec:method:str}.

% -------------------------------------------------------------------
\subsection{Unsupervised learning: Mixture model for foehn classification}\label{sec:method:mixture}

As direct foehn measurement do not exist, the data from the AWSs are currently
unlabeled. Therefore, a mixture model is employed to distinguish between the
`foehn' and `no foehn' events:
\begin{equation}
    \text{Pr}_{\text{obs}}(\text{foehn}) = f_{\text{mix}}(\text{measurements}).
    \label{eqn:foehnevent}
\end{equation}
$\text{Pr}_{\text{obs}}(\text{foehn})$ denotes the posterior
probability for a `foehn' observation at a specific time and station, which is
modeled as a function ($f_{\text{mix}}()$) of the 10\,min in-situ measurements
(Sec.\,\ref{sec:data:obs}).
We employ a two-component Gaussian mixture model \citep{gruen2008}
with concomitants, closely following the method proposed by \cite{plavcan2014}
implemented in the \emph{R}~package \verb|foehnix| \citep{staufferfoehnix}.

The prerequisite condition for an observation to be used for classification is
that the wind direction at the location of interest falls within the
prevailing foehn direction at the target location, and that wind from a
specific direction is also prevalent at the corresponding crest station at the
same time (details in Tab.\,\ref{app:tab:foehnixconfig}). Only the periods
matching this precondition are used for estimating the Gaussian mixture model,
while $\text{Pr}_{\text{obs}}(\text{foehn}) = 0$ is set for all
remaining observations.

The underlying concept involves that two unobservable Gaussian components (or
clusters) exist, one describing `foehn' conditions and the other `no foehn'
conditions. To distinguish between these two components, a main covariate is
required. In this study, the potential temperature difference ($\Delta \theta$)
between the valley station ($t_{\text{valley}}$) and the crest station
($t_{\text{crest}}$) is used, employing a simple
dry adiabatic temperature reduction $\Delta \theta = t_{\text{valley}} -
t_{\text{crest}} - 0.01 \Delta h$ based on the difference in altitude ($\Delta
h$; cf.Tab.\,\ref{tab:stations}\,\&\,\ref{app:tab:foehnixconfig}).
During foehn events, the air descends on the leeward side of the mountains,
resulting in a well-mixed atmosphere with neutral stratification, hence
a potential temperature difference close to zero.

\begin{figure}[!ht]
    \center
    \includegraphics[width=.5\textwidth]{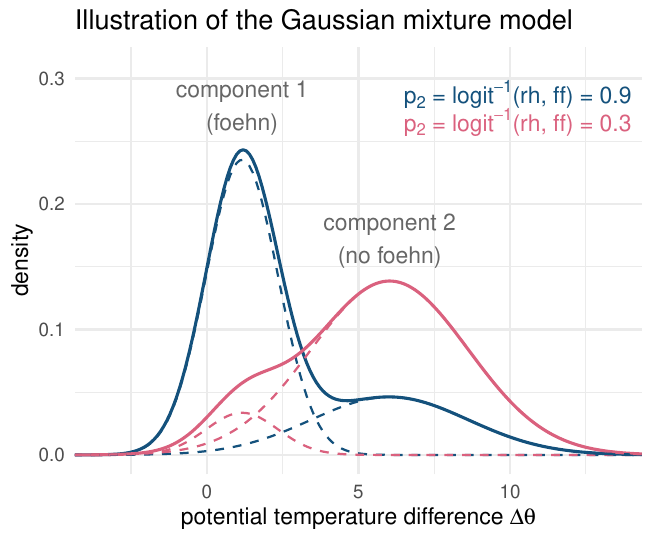}
    \caption{Illustration of the two-component Gaussian mixture model used for classification.
    The potential temperature difference ($\Delta \theta$) is used as the main variable to separate component 1 (foehn) and component 2 (no foehn);
    the concomitant model includes relative humidity (rh) and mean wind speed (ff) at the valley station.
    Solid lines: density of the mixture model once with a probability $p_2 = 0.9$
    to observe component 2, once with $p_2 = 0.3$.}
    \label{fig:mixturemodel}
\end{figure}

However, the potential temperature difference alone might not be sufficient to
adequately separate the two states. Therefore, an additional concomitant model
is employed to weigh the two components conditional on additional covariates.
In this study, binary logistic regression is used for the concomitant model,
employing relative humidity and mean wind speed as additional covariates. Models with this
specification have been shown to work well for stations in the Alpine region
(e.g.,\ \citealt{plavcan2014,plavcan2015}). Figure\,\ref{fig:mixturemodel}
provides an illustration of this model, depicting the use of $\Delta \theta$ to
separate the two components (`foehn'/`no foehn') and the effect of the
concomitant model on the joint density. More details can be
found in Appendix\,\ref{app:sec:method:mixture}.

Once estimated, the mixture model provides the posterior probability
$\text{Pr}_{\text{obs}}(\text{foehn})$ for each 10\,min interval
where the required in-situ measurements are available (Eqn.\,\ref{eqn:foehnevent}).
As these foehn `observations' need to be combined
with ERA5 in the next step, the results are upscaled
to an hourly temporal resolution ($\text{Pr}_{\text{1h}}$).
%%%to match the temporal resolution of the reanalysis.
Therefore, the following basic assumption is applied: Each hour is considered a
`foehn' event if at least half of the 10\,min posterior probabilities
within that hour (e.g., 00:10--01:00\,UTC for 01:00\,UTC) are $\ge 0.5$, otherwise
the hour is considered a `no foehn' event.
If fewer than 4 out of 6 individual 10\,min probabilities are available, the hour
is excluded. This is similar to \citet{gutermann2012} and \citet{mony2021}
who employ a `four out of six rule'.
\begin{equation}
    \text{Pr}_{\text{1h}} = \begin{cases}
        \text{missing}  & \text{if} ~~~ \sum \text{Pr}_{\text{obs}} \in [0-1] < 4 \\
        \text{foehn} & \text{if} ~~~ \frac{1}{N} \sum \big(\text{Pr}_{\text{obs}} \ge 0.5 \big) \ge 0.5 \\
        \text{no foehn} & \text{else}
    \end{cases}
    \label{eqn:method:hourly}
\end{equation}

% -------------------------------------------------------------------
\subsection{Supervised learning: Modeling foehn probability}\label{sec:method:glm}

The outcome of the previous section is a binary time series,
with each hour labeled as either a `foehn' or `no foehn' observation 
(Eqn.\,\ref{eqn:method:hourly}). This serves as the response variable in the
supervised machine learning model, using the ERA5 data (Sec.\,\ref{sec:data:era5})
as input variables. The goal is to capture the relationship between the two data sets
with model of the form
\begin{equation}
    \text{Pr}_{\text{1h}}(\text{foehn}) = f_{\text{learn}}(\text{ERA5}),
    \label{eqn:supervised}
\end{equation}
where the probability $\text{Pr}_{\text{1h}}$ of the binary response is
modeled as a function $f_{\text{learn}}()$ of the available information
extracted from ERA5.
$f_{\text{learn}}()$ can be any learner suitable for a binary response such as
logistic regression, decision trees, random forests, or neural
networks, to mention a few. For this study, three different learners/models
are employed:

\begin{description}
    \item[lasso] Logistic regression with lasso (L1) regularization \citep{friedman2008,tay2023}.
    \item[stabsel] Logistic regression with lasso-based stability selection \citep{meinhausen2010}.
    \item[xgboost] Extreme gradient boosting \citep{chen2016}.
\end{description}

In order to investigate the possible benefits of incorporating large-scale
information from the stations' neighborhood, two variations of each learner are
considered: one utilizing the `full' set of 497 variables and one only using
the `direct' set of 155 variables (Sec.\,\ref{sec:data:era5}).

To account for location and time of day, separate models are estimated for each
of the six stations (Tab.\,\ref{app:tab:foehnixconfig}) for each hour of the day
($0000$\,UTC, $0100$\,UTC, \dots, $2300$\,UTC), resulting in a total of
864\,models. Depending on the station, the training data for these models
include 10--18 years of data (see Tab.\,\ref{tab:stations}).
In addition, a six-fold cross-validation (CV) is performed using a fixed period
of 12 years (2011--2022) where, in each fold, two consecutive years are left
out as test data. Ellb\"ogen and Innsbruck are missing one fold (with test
data 2013--2014) where no measurements from the crest
station are available and thus the classification is not possible.

% -------------------------------------------------------------------
\subsection{Long-term foehn reconstruction}\label{sec:method:reconstruction}

Once the models from the previous section are estimated, they can be applied to
the entire ERA5 period available. Although this is a prediction from a
statistical perspective, it is termed `reconstruction' in this
article, as these predictions are applied backwards in time. The result is an
hourly time series of foehn probabilities $\widehat{\text{Pr}}_{\text{1h}}$
from January 1, 1940 to December 31, 2022 (83 years).

This reconstruction can fill possible gaps in historic records during the lifetime
of an AWS and extend `foehn observations' to times prior to or beyond the installation period 
of the AWS. The results may also serve as valuable input for investigating
the effects of foehn, e.g., in ecology where the warming and drying
effects of frequent foehn events can significantly impact flora and fauna
or favor the occurrence of forest fires.

Moreover, the reconstruction offers the possibility to analyze foehn
occurrence from a climatological perspective: Did the occurrence of foehn
increase/decrease along with the changing climate over the decades?
Are there changes in the seasonal or diurnal patterns? These questions are
investigated in more detail in the next section.

% -------------------------------------------------------------------
\subsection{Season-trend decomposition}\label{sec:method:str}

The comprehensive reconstructed data set allows to
study foehn occurrence in a climatological context.
%to identify possible trends or anomalies over the past decades.
For this analysis, the hourly probability
(Eqn.\,\ref{eqn:supervised}) is aggregated by (i) taking the highest
probability $\widehat{\text{Pr}}_{\text{1h}}$ per day (0000\,UTC--0000\,UTC), before (ii) calculating
monthly averages. The resulting time series contains ``monthly means of the
daily maxima'', which are then modeled using a season-trend decomposition.

Due to the nature of the data there is a large year-by-year but also
within-year variability depending on the prevailing weather situation. To
decompose the signal, a season-trend decomposition is employed
separating the signal into long-term changes and a remainder component
containing the residual variability.

In this study, the regression-based decomposition of \citet{dokumentov2022} is used,
which also provides confidence intervals for the estimated season and trend components.
The model for the monthly mean foehn probabilities $y_t$ assumes an additive
decomposition into a smoothly changing long-term trend $T_t$, a smoothly
changing seasonal component $S_t^{(m)}$, and a remainder $R_t$:
\begin{equation}
    y_t = T_t + \sum_{m = 1}^{12} S_t^{(m)} + R_t~~\text{with}~~t \in 1, \dots, J,
    \label{eqn:method:str}
\end{equation}
where $m \in \{1, 12\}$ is the seasonal frequency (i.e., monthly) and $J$ is
the sample size (83 years $\times$ 12 months = 996). The model is estimated
via the \emph{R}~package \verb|stR| \citep{dokumentovstR}.
%%%The trend-season model is estimated using five-fold cross-validation to account for temporal
%%%autocorrelation using the \emph{R}\,package \verb|stR| \citep{dokumentovstR}.

% Note to Reto: Only works with 5-fold CV, I've testes nFold = 20 with
% a custom method (adjusted AutoSTR) with only minor differences..
%
% 12 * 2 is two years (try gapCV = 1000) but can have strong impact
% on the result, gapCV = 3 * 12 for Ellboegen seems to kill the ternd signal,
% with four years it looks similar to two years again.
%
% AutoSTR(xts, robust = FALSE, gapCV = 12 * 2, lambdas = NULL,
%         reltol = 0.001, confidence = c(0.95),  nsKnots = NULL (12),
%         trace = FALSE)

%% ========================================================================
%% ========================================================================
%% ========================================================================
\section{Results}

First, this section investigates which insights can be gained from the
reconstruction about the foehn occurrence at the six different target stations.
Different temporal scales are considered for this, namely:
Inter-annual changes in the foehn probabilities in Section\,\ref{sec:res:annual},
long-term trends and seasonal patterns in Section\,\ref{sec:res:str}, as well as
changes in the diurnal patterns across decades in Section\,\ref{sec:res:clim}.
All of these results are based on the reconstruction using the `lasso' learner
with the `full' covariate set for the full time period (without cross-validation).

Second, the performance of the supervised learning model is assessed under
different model specifications, namely: Using the `full' set of all 497 variables
vs.\ the 155 `direct' variables only in Section\,\ref{sec:res:fullset} and
comparing the performance of the three supervised learners (lasso, stability
selection, extreme gradient boosting) in Section\,\ref{sec:res:comparison}.
All of these results are based on out-of-sample Brier scores obtained in a
six-fold cross-validation.

%% ------------------------------------------------------------------------
\subsection{Temporal changes: Average annual foehn probabilities}\label{sec:res:annual}

The primary outcome of this study is the complete reconstruction of hourly foehn
probabilities over $83$\,years (see Sections\,\ref{sec:method:mixture}--\ref{sec:method:reconstruction}),
yielding time series with $N\sim7.27\cdot10^5$ observations. To carve out
inter-annual variations in the foehn probabilities, we aggregate the reconstructed
data by taking the daily maximum of $\widehat{\text{Pr}}_{\text{1h}}$ (Eqn.\,\ref{eqn:supervised})
before calculating annual means. This can be interpreted as the average probability
of observing a foehn event on any given day within that year.

\begin{figure}[!ht]\centering
    \includegraphics[width=\textwidth, page=5]{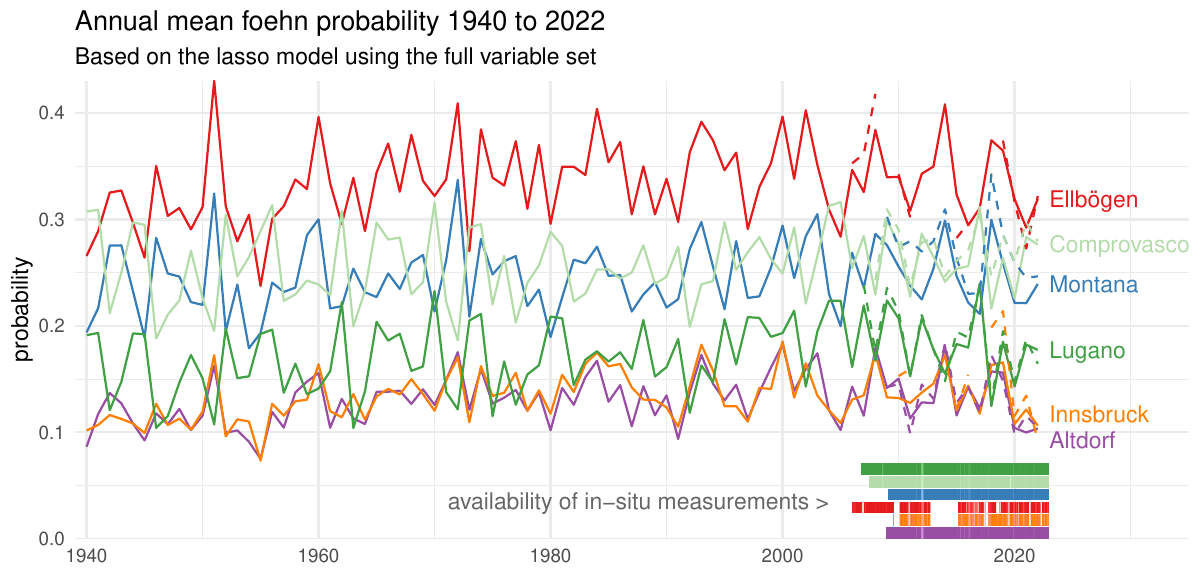}
    \caption{\label{fig:res:annual}
        Annual mean of highest daily foehn probability (solid lines) for all
        six stations from 1940 to 2022 based on the `lasso' model with `full' covariate
        set. Additionally, annual means of daily maxima from the foehn classification using AWS data
        (dashed lines) and the availability of in-situ measurements (straight bands at the bottom) are shown.
        Appendix\,\ref{app:sec:reconstruction} shows a comparison to all other models and variants.
    }
\end{figure}

Figure\,\ref{fig:res:annual} contains the result for all six stations, with
Ellb\"ogen exhibiting the highest mean annual probabilities (on average
$0.334$), while Altdorf and Innsbruck
show the lowest (on average $0.130$ and
$0.133$ respectively). Additionally, the
annual mean of daily maxima from the classification is shown for years with at
least 80\% of measurements available at the AWSs.
The results show an overall good agreement between the two signals from the
reconstruction and from the classification, with some larger gaps due to data
availability as well as some noticeable differences for specific stations in
particular years.

The reconstruction reveals a pronounced inter-annual variability, with certain years
exhibiting a much higher annual mean than the long-term average, whilst others
distinctly fall below.
This variability is not random, as one can see similar patterns among the four
south-foehn stations. For instance, all four stations show unusually high mean probabilities
for 1951 and 1972. Similarly, the two north-foehn stations exhibit a similar
temporal behavior over time.

Moreover, the figure suggests a possible increase in the annual mean foehn
probability for south-foehn stations between 1940 and 1980. Hence this question
is investigated in more detail in the next section.

%% ------------------------------------------------------------------------
\subsection{Temporal changes: Climatological trends and seasonal patterns}\label{sec:res:str}

\begin{figure}[!ht]
    \includegraphics[width=\textwidth]{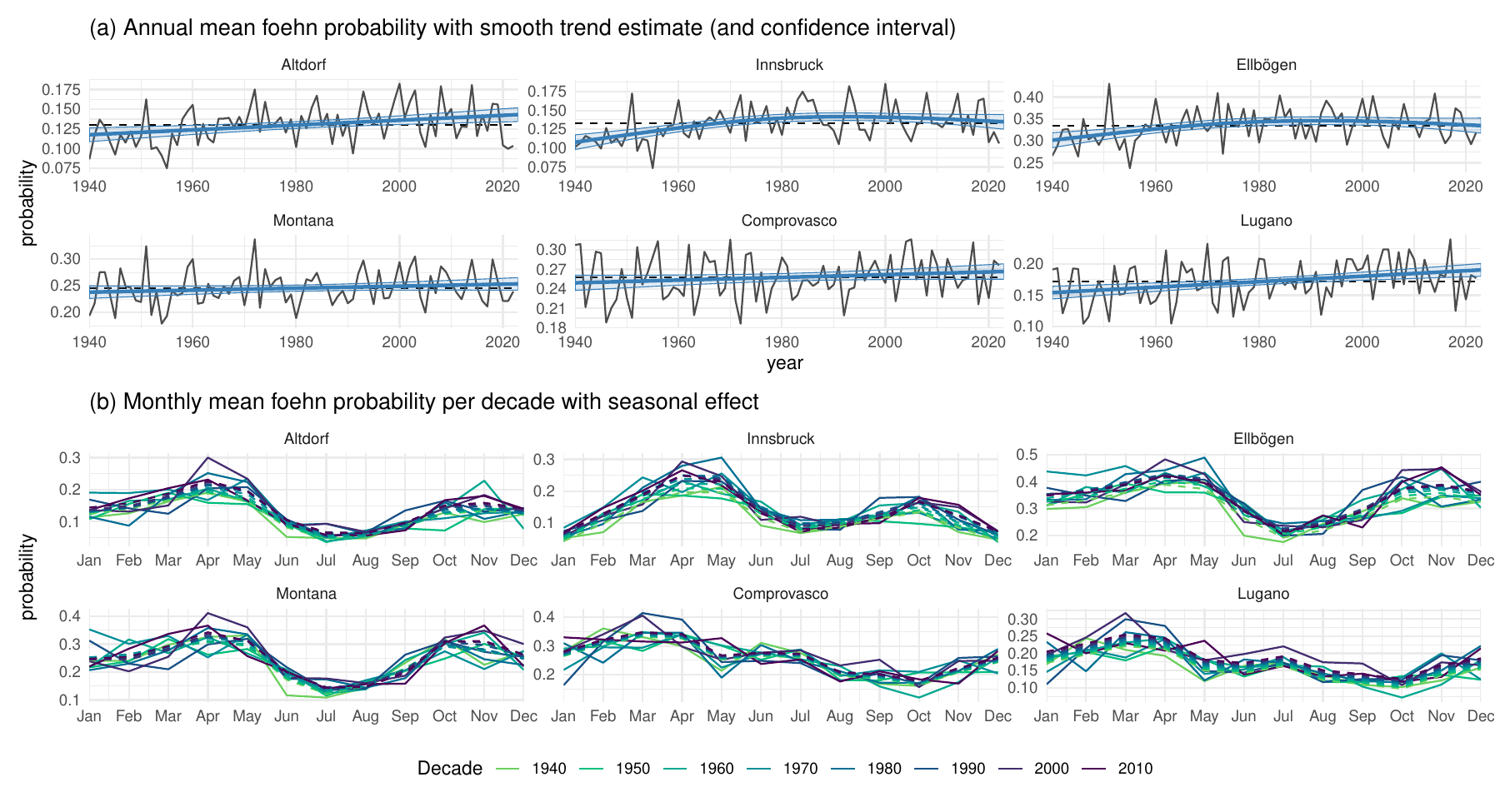}
    \caption{\label{fig:res:str}
        Results of the season-trend decomposition. (a) Estimated long-term
        trends with estimated 95\% confidence interval (blue) on top of annual mean probabilities
        (black), the horizontal dashed line shows the mean of the trend for better orientation.
        (b) Estimated seasonal pattern averaged over the whole period 1940--2022 (dashed) 
        and separately by decade (solid, label indicates first year of each decade).
    }
\end{figure}

In this section, the analysis from the previous section is taken a step further.
Rather than focusing on the inter-annual variation the goal is to bring out
the long-term climatological trends and changes in the seasonal patterns.
Hence the reconstructed hourly time series are again aggregated but to monthly
(rather than annual) means of the daily maxima of $\widehat{\text{Pr}}_{\text{1h}}$.
Based on the season-trend decomposition outlined in Section\,\ref{sec:method}
Figure\,\ref{fig:res:str}(a) illustrates the resulting smooth trends ($T_t$)
along with the corresponding $95\%$ confidence intervals. Figure\,\ref{fig:res:str}(b)
depicts the corresponding smoothly varying seasonal signals ($S_t^{(m)}$) averaged
over decades for visual purposes (where decade 1940 corresponds to the years
1940--1949 etc.).

The estimated trends (Fig.\,\ref{fig:res:str}a) show a significant
increase for the two stations in Western Austria (Innsbruck, Ellb\"ogen) between
1940 and 1980, and a plateau thereafter. All other stations show a linear
increase over the study period. For four of the six stations (all except
Montana and Comprovasco) the smooth trend differs significantly from
a constant.

% Notes reto (this is with gapCV = 12 and nFold = 16; custom AutoSTR adaption).
%                Significant      Increase            Decrease
% Altdorf         barely not      Apr, Mai, Oct       Feb, Aug, Dez
% Ellboegen       no              Oct, Nov, Dec       Jul, Aug, Sep
% InnsbruckUNI    spring          Apr, Oct            Jul, Aug, Sep
% Montana         no              Apr, Oct, Nov       Aug, Sep
% --- do not show anything ---
% Comprovasco     nothing
% Lugano          nothing

Figure\,\ref{fig:res:str}(b) shows the analysis of the seasonal changes
which reveals the different characteristics between
north-foehn stations and south-foehn stations. The two stations located south
of the main Alpine ridge, Comprovasco and Lugano, show one distinct maximum in
spring and a minimum during autumn. This pattern is stable over the entire
study period and no changes in the seasonal pattern ($S_t^{(m)}$) are found.
The picture looks different when focusing on the four south-foehn stations which
all show two maxima in spring and autumn with lower probability of foehn
occurrence during summer and winter. Although not significant, the season-trend
decomposition indicates an increase in the probability of foehn in spring
(April, Mai) as well as in autumn (October, November) with a slight decrease in
late summer (August, September).

%% ------------------------------------------------------------------------
\subsection{Temporal changes: Intra-daily patterns}\label{sec:res:clim}

For certain applications, information about intra-daily patterns
and their changes over time can be of great interest. With the hourly temporal
resolution of the reconstruction, such insights are now possible across
several decades.

\begin{figure}[!ht]\centering
    \includegraphics[width=\textwidth, page=3]{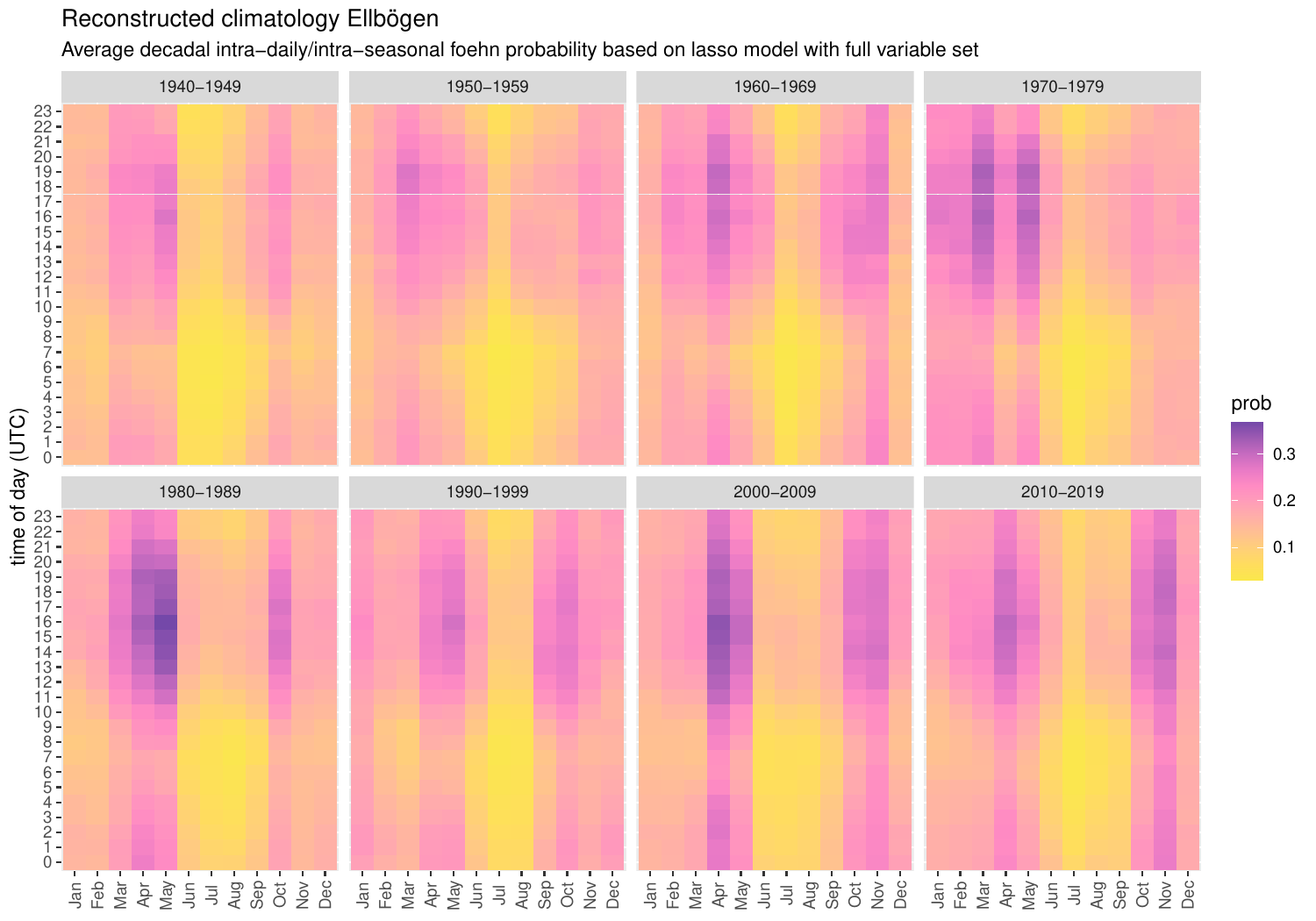}
    \caption{\label{fig:res:clim}
        Hovm{\o}ller diagrams showing the average foehn probability for
        Ellb\"ogen over the course of the day (y-axis) for all months (x-axis)
        for the decades 1940--2019.
    }
\end{figure}

Figure\,\ref{fig:res:clim} shows Hovm{\o}ller diagrams for Ellb\"ogen,
depicting the decadal mean probability per time of day and month.
Despite pronounced variability between the decades, the plot supports
the previous findings showing an overall increase over time with strongest
increase in spring (April, May) and in autumn (October, November).
In addition, this visualization gives insights into the
diurnal pattern. Generally, foehn occurrence in Ellb\"ogen is more likely
during the day (around 1000\,UTC--2200\,UTC) in spring and autumn, the period
where indication for a certain increase was found (Sec.\,\ref{sec:res:str}).
The minimum average foehn probability is in the early morning and tied to the length of the
night. The time of the minimum varies somewhat interdecadally.

%% ------------------------------------------------------------------------
\subsection{Model performance: Benefit of full covariate set}\label{sec:res:fullset}

As described in Section\,\ref{sec:method:glm}, two variants of the supervised
learning methods are estimated: one using only the 155 `direct' variables and one
using the 497 `full' variables including large-scale atmospheric conditions
(Sec.\,\ref{sec:data:era5}). For both variants, a six-fold CV is performed
(Sec.\,\ref{sec:method:glm}).

\begin{figure}[!ht]
    \includegraphics[width=\textwidth]{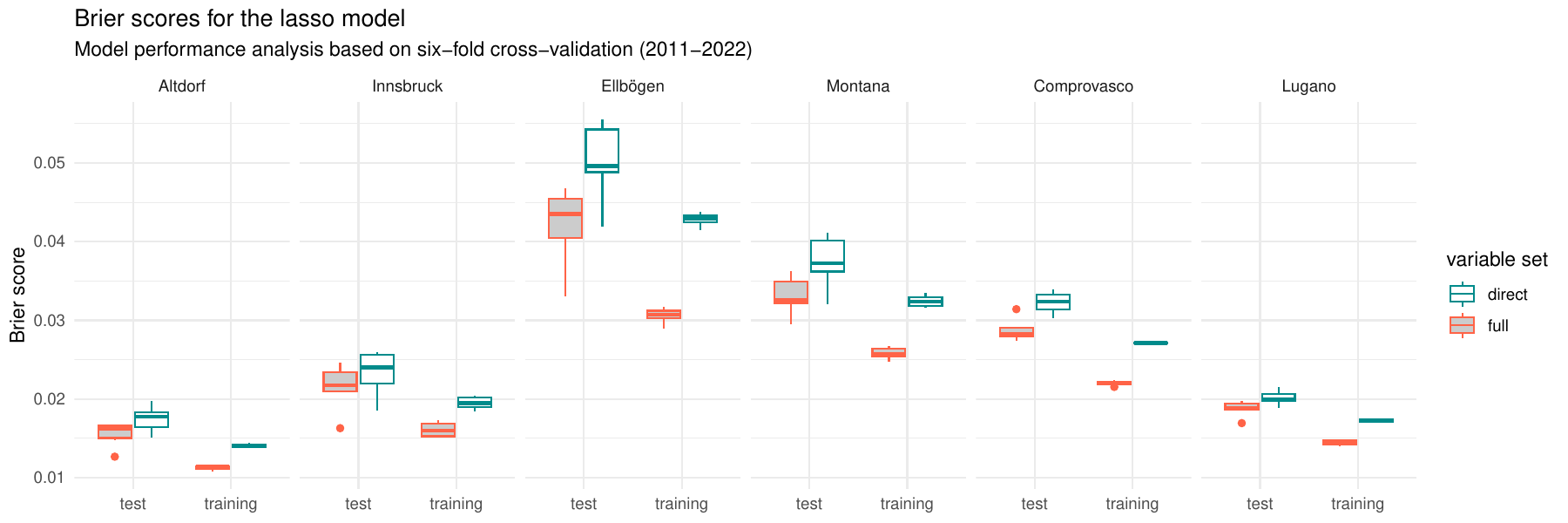}
    \caption{\label{fig:res:brier}
        Comparison of `lasso' model performance using the `direct' variable set
        (blueish) versus the larger `full' variable set (orange with gray filling)
        for all stations. Shown is the average Brier score from the six-fold
        CV for the test (out-of-sample) and training (ins-sample)
        period. Lower is better.
    }
\end{figure}

To investigate the benefit of the `full' covariate set,
Figure\,\ref{fig:res:brier} shows the Brier scores (BSs) for the `lasso' model
for all six stations. For each station and each variant, BSs are shown for the
test data set (out-of-sample) as well as for the training data set (in-sample).
This shows that the models based on the `full' variable set
clearly outperform the models based on the `direct' variables only.
Although the overall performance of the less complex models based on the
`direct' variable set is still decent, including the additional large-scale
spatio-temporal information substantially improves the overall model performance.

In addition to the predictive skill, Figure\,\ref{fig:res:brier} also shows
the stability of the model with largest variance in the BSs visible
on the test data sets in Ellb\"ogen. On the training data sets the scores
barely vary due to the large sample size (10 years, hourly data).

%% ------------------------------------------------------------------------
\subsection{Model performance: Comparison of supervised learners}\label{sec:res:comparison}

While the previous section focuses on the benefits of using more input data,
this section compares the three different supervised learners described in
Section\,\ref{sec:method:glm}. For simplicity, only the results for models based on the `full'
variable set are shown as they have been shown to outperform those only using the
`direct' variable set (Sec.\,\ref{sec:res:fullset}).

\begin{figure}[!ht]
    \includegraphics[width=\textwidth]{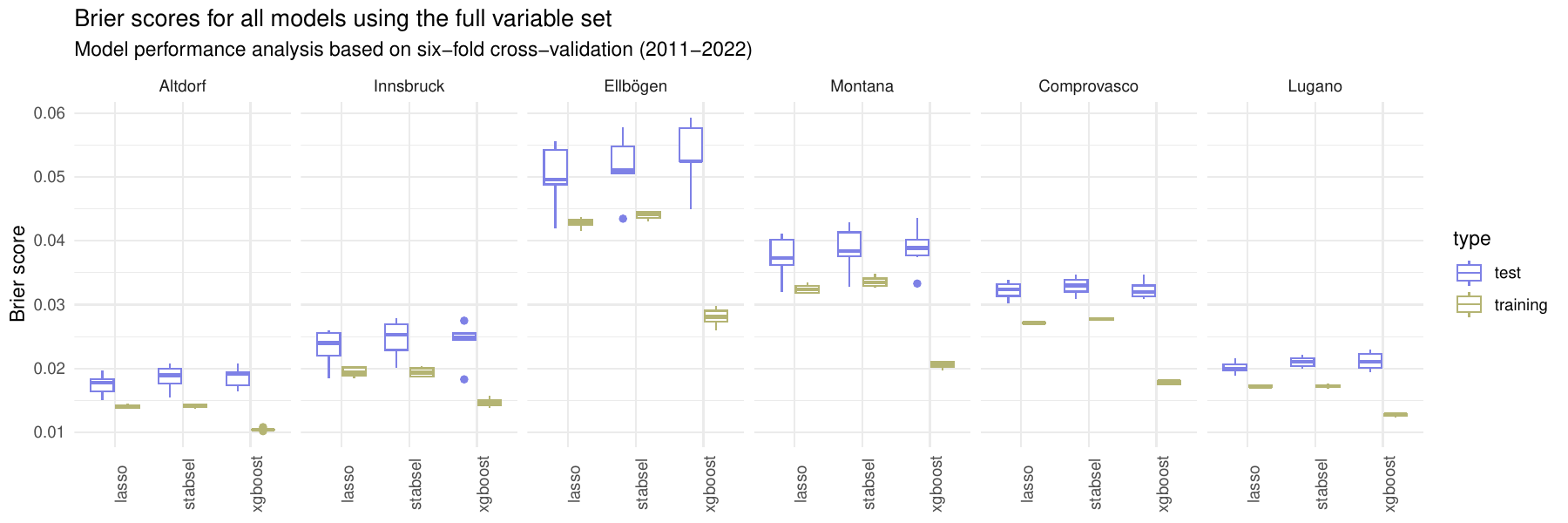}
    \caption{\label{fig:res:brierallmodels}
        Comparison of all supervised learning models using the `full' variable
        set. Brier scores for training (in-sample; green) and test (out-of-sample; violet)
        data set based on six-fold cross-validation. Lower is better.
    }
\end{figure}

Figure\,\ref{fig:res:brierallmodels} illustrates that the BSs from the six-fold
CV on the test set (out-of-sample) is comparable for all three supervised learners
with only minor differences. The average BS is slightly lower for
`lasso' ($0.0261$),
followed by `stabsel' ($0.0276$),
and `xgboost' ($0.0283$).

On the training data (in-sample) the picture is similar for `lasso' and `stabsel'
but `xgboost' has much lower BSs. This indicates that `xgboost' might be subject to
some overfitting, despite careful tuning of the hyperparameters (App.\,\ref{app:model:xgboost}).

\section{Discussion and outlook}\label{sec:discussion}

% motivation
Using a novel combination of unsupervised and supervised learning we are able
to accurately reconstruct long-term foehn time series (starting from 1940) at
hourly resolution. More specifically, foehn classification is first accomplished by
(unsupervised) Gaussian finite mixture models based on AWS measurements and then
linked to ERA5 data using binary supervised learners such as lasso, stability
selection, or extreme gradient boosting. The resulting foehn reconstruction
enables novel analyses, exemplified here by investigating long-term changes
in trends, seasonal patterns, and diurnal cycles of foehn occurrence.

% discuss results (recon + trends)
The season-trend decomposition based on the period 1940--2022 reveals that
all six stations considered have either experienced a linear increase in
foehn occurrence (probability) over the entire study period or an increase
between 1940 and the early 1980s that leveled off afterwards. For four of the
six stations it can be shown that this increase is significant.
Although no significant change in the seasonality is found, the results for
all south-foehn stations (Altdorf, Montana, Ellb\"ogen, and Innsbruck)
indicate a slight increase in the occurrence of foehn in spring and autumn,
with a slight decrease in late summer.

The high quality of the foehn reconstruction is partially due to using a
large set of predictor covariates that not only contains information at the target
location but also includes additional large-scale atmospheric information
from the stations' surrounding. The benefits of this full set of covariates
are similar for all three supervised learners considered: logistic regression
with lasso regularization (`lasso'), logistic-regression-based stability
selection (`stabsel'), and extreme gradient boosting (`xgboost'). Lasso
performs best in our application, closely followed by the other two learners.
Some further improvements might be gained for xgboost if overfitting on the
training data can be further reduced, e.g., by a different hyperparameter
tuning strategy.

For a comparison to existing publications, we complement the Brier scores
shown in the main paper by other popular scores for binary outcomes, such as the
false negative rate (FNR, also known as miss rate),
false positive rate (FPR, also known as false alarm rate), and
percent correct (PC, also known as accuracy).
Based on the best model (lasso with full covariate set), we obtain the
following performances for Altdorf: $15.7\%$/$0.4\%$/$98.8\%$ (FNR/FPR/PC).
These align well with the existing literature and stand out for an exceptionally low FPR.
\cite{sprenger2017} report $11.8\%$/$33.8\%$/$96.5\%$ and \cite{mony2021}
report $21.4\%$/$21.4\%$/$97.7\%$. Similarly for Lugano we obtain
$15.3\%$/$0.7\%$/$98.2\%$, while \cite{mony2021} report $22.1\%$/$22.1\%$/$97.1\%$.
More details are included in the appendix (App.\,\ref{app:sec:modelperformance}).

%compare to swiss foehn
The same holds for the foehn classification when compared to existing literature
and the Swiss foehn index (SFI) operationally used at MeteoSwiss in terms of
``average foehn hours per year''.
The results from the Gaussian mixture model (Sec.\,\ref{sec:method:mixture})
for Altdorf show an average of 482.4\,h/year which aligns well with the SFI
(458.4\,h/year), as well as the results reported by \cite{duerr2008}
(478\,h/year), \cite{jansing2022} (465.8\,h/year), and \cite{ms2024}
(477\,h/year).
Montana exhibits 1007.4\,h/year (SFI 904.7\,h/year\footnote{Average based on
the years 2009--2016 only.}),
%, \cite{ms2024} 542\,h/year\footnote{TODO(R): This number looks fishy, see footnote SFI}),
while Lugano and Comprovasco show
644.7\,h/year (SFI 563.0\,h/year, \cite{ms2024} 551\,h/year) and 1077.4\,h/year (SFI
953.4\,h/year), respectively.
%The two Austrian stations show an average of
%1529.5\,h/year for Ellb\"ogen and 362.8\,h/year for Innsbruck.

It should be emphasized that the excellent performance of the combined approach
depends crucially on the high quality of the automatic foehn classification in
the Gaussian finite mixture model. Thus, if the separation of the AWS measurements
into `foehn' or `no foehn' components works well (as for the six stations presented),
then different binary supervised learners are able to link this reliably to
the ERA5 data. However, there may also be stations where a distinction into just
two components might not be sufficient. For example, we have found this to be
the case for Aigle, Switzerland, where a three-component mixture model appears to
be necessary to separate light down-valley winds, strong humid (catabatic)
outflows, and actual foehn situations (details not shown).

%% It would be interesting
%% in the future to extend the binary approach introduced in this paper to a multinomial
%% approach. Thus, the outcome from a three-component finite mixture model could
%% then be linked to ERA5 data with supervised learners that can deal with three
%% response categories. The same idea could be applied when using a classification,
%% as employed operationally at MeteoSwiss, distinguishing `foehn`, `no foehn`, and
%% `mixed foehn conditions'.
%% 
%% \emph{FIXME: Maybe conclude with some remark that the actual reconstructed time
%% series, at least for Innsbruck and Ellb\"ogen, are available for download and that
%% we hope that these can be utilized as inputs in other studies?}

%% -- Optional special unnumbered sections -------------------------------------

\section*{Computational details}

The results in this paper were obtained using \proglang{R}~4.2+.
The majority of data preparation and handling is done using the
\proglang{R} packages \pkg{stars}~0.6.4, \pkg{sf}~1.0.15 and \pkg{zoo}~1.8.12.
\pkg{foehnix}~0.1.6 is used for foehn classification, the supervised learning
is based on \pkg{glmnet}~4.1.7 and \pkg{xgboost}~1.7.5.1. The season-trend
decomposition is based on the \proglang{R} package \pkg{stR}~0.6.

\section*{Acknowledgments}

The study is partly based on the preliminary work of \cite{morgenstern2020}.
The computational results presented here have been achieved (in part) using the
LEO HPC infrastructure of the University of Innsbruck.

%% -- Bibliography -------------------------------------------------------------
\bibliography{references}

%% -- Appendix (if any) --------------------------------------------------------
\newpage

\begin{appendix}

\newpage

%% ====================================================================
%% ====================================================================
\section{ERA5 fields}\label{app:sec:data:era5}

As described in Section\,\ref{sec:data:era5} information from 90 different
ERA5 fields is bilinearely interpolated to (i) the target location of the
different stations used (cf.\,\ref{sec:data:obs}) as well as to a series
of neighboring points arranged in a `star' formation to include large-scale
information. The fields used are listed in Table\,\ref{app:tab:era5:sf} and
\ref{app:tab:era5:pl}.

\begin{table}[th]
    \begin{footnotesize}
        \begin{tabular}{l l}
        \multicolumn{2}{l}{CDS ERA5 fields name} \\
           \hline
    \verb|100m_u_component_of_wind| & \verb|100m_v_component_of_wind| \\
    \verb|10m_u_component_of_wind| & \verb|10m_v_component_of_wind| \\
    \verb|10m_wind_gust_since_previous_post_processing| & \verb|instantaneous_10m_wind_gust| \\
    \verb|2m_dewpoint_temperature| & \verb|2m_temperature| \\
    \verb|surface_pressure| & \verb|mean_sea_level_pressure| \\
    \verb|eastward_gravity_wave_surface_stress| & \verb|northward_gravity_wave_surface_stress| \\
    \verb|convective_precipitation| & \verb|large_scale_precipitation| \\
    \verb|total_precipitation| & \verb|boundary_layer_dissipation| \\
    \verb|boundary_layer_height| & \verb|charnock| \\
    \verb|friction_velocity| & \verb|gravity_wave_dissipation| \\
    \verb|high_cloud_cover| & \verb|low_cloud_cover| \\
    \verb|medium_cloud_cover| & \verb|total_cloud_cover| \\
    \verb|total_column_cloud_liquid_water| & \verb|surface_net_solar_radiation| \\
    \verb|surface_net_thermal_radiation| & \verb|surface_sensible_heat_flux| \\
    \verb|surface_solar_radiation_downwards| & \verb|surface_thermal_radiation_downwards| \\
    \hline
        \end{tabular}
    \end{footnotesize}
        \caption{List of ERA5 single level field retrieved.}
        \label{app:tab:era5:sf}
\end{table}

\begin{table}[h]
    \begin{footnotesize}
    \begin{tabular}{l l}
        \multicolumn{2}{l}{CDS ERA5 fields name} \\
           \hline
           \verb|divergence| & \verb|geopotential| \\
           \verb|potential_vorticity| & \verb|specific_humidity| \\
           \verb|temperature| & \verb|u_component_of_wind| \\
           \verb|v_component_of_wind| & \verb|vertical_velocity| \\
           \verb|specific_cloud_liquid_water_content| & \verb|vorticity| \\
           \hline
    \end{tabular}
    \end{footnotesize}
    \caption{List of ERA5 pressure level fields retrieved; each on the
        standard pressure levels 500\,hPa, 700\,hPa, 750\,hPa, 800\,hPa, 850\,hPa and 900\,hPa.}
        \label{app:tab:era5:pl}
\end{table}

To keep the approach generic, these neighbouring points are `arbitrary' points
not linked to stations or cities, but geometrically placed around the
target station relative to its location to the main mountain range.
Figure\,\ref{fig:neighbors} (main article) shows these `stars' which
are constructed as follows: First, the
orthogonal from the center location (station location; \verb|C|) to the
main mountain range is computed. Along this orthogonal axis, two additional
points upwind of the foehn wind direction (\verb|U|, \verb|UU|) and two
points in the downwind direction (\verb|D|, \verb|DD|) are defined with a
radius of 1$^\circ$ and 2$^\circ$ from the center point. Second, four
points with a radius of 2$^\circ$ from the center point deviating by
45$^\circ$ from the orthogonal are added, defining points on the right hand
side (\verb|*R|) and left hand side (\verb|*L|) of the main foehn direction
in the upwind direction (\verb|UR|, \verb|UL|) and downwind direction
(\verb|DR|, \verb|DL|). The downwind direction for stations located north
of the main alpine range is towards north (south-foehn stations), for those
located south of the main alpine range the downwind direction is toward
south (north-foehn stations).

Based on these points, the `direct' variable set (155 variables) as well
as the `full' variable set (497 variables) are calculated as described
in Section\,\ref{sec:data:era5} (main article). 
The extended (`full') variable set contains a series of spatial and
spatio-temporal information such as:

\begin{itemize}
    \item surface pressure differences between
        \verb|C| to \verb|U|, \verb|U| to \verb|C|, \verb|UL| to \verb|DR| and others,
    \item potential temperature differences between
        \verb|C| to \verb|U|, \verb|U| to \verb|C|, \verb|UL| to \verb|DR| and others,
    \item differences in vertical temperature differences between
        \verb|C| to \verb|U|, \verb|U| to \verb|C|, \verb|UL| to \verb|DR| and others,
    \item sum of precipitation on the upwind side (sum of \verb|U|, \verb|UU|, \verb|UR|, \verb|UL|)
        as well as on the downwind side (sum of \verb|D|, \verb|DD|, \verb|DR|, \verb|DL|) as well
        and the difference between these two sums,
    \item mean cloud cover on the upwind side (sum of \verb|U|, \verb|UU|, \verb|UR|, \verb|UL|)
        as well as on the downwind side (sum of \verb|D|, \verb|DD|, \verb|DR|, \verb|DL|) as well
        and the difference between these two means.
    %%%%\item \dots
\end{itemize}

Moreover, temporal variables are calculated including e.g., changes in the
geopotential height, temperature, or changes in humidity on different levels over the past
3 hours or over the next 3 hours and variations thereof.

%%%%TODO(R): There is also a series of time-aggregated variables
%%%%(see `frecon/R/prepare.R`).
%%%%
%%%%DISABLED; MAKES NO SENSE THIS WAY
%%%%
%%%%\begin{footnotesize}
%%%%<<r, echo = FALSE, results = tex>>=
%%%%library("frecon")
%%%%locs <- c("DD", "DR", "DL", "D", "C", "U", "UL", "UR", "UU")
%%%%tmp <- frecon:::fr_get_dobjs(locs)
%%%%tmp <- data.frame("description" = sapply(tmp, function(x) x$desc),
%%%%                  "formula"     = sapply(tmp, function(x) x$formula))
%%%%tmp$formula <- paste("........", gsub("\\s+", " ", tmp$formula))
%%%%
%%%%
%%%%tmp <- data.frame("Derived variables" = c(rbind(tmp$description, tmp$formula)))
%%%%
%%%%library("xtable")
%%%%x.big <- xtable(tmp, label = 'tabbig', align = c("l", "p{15cm}"),
%%%%                caption = 'Example of longtable spanning several pages')
%%%%print(x.big, tabular.environment = 'longtable', floating = FALSE,
%%%%      include.rownames=FALSE)
%%%%@
%%%%\end{footnotesize}

%% ====================================================================
%% ====================================================================
\section{Unsupervised learning: Two-component mixture model}\label{app:sec:method:mixture}

For all eight AWSs used (Tab.~\ref{tab:stations}; main article)
observations are available for the most recent 10--18 years on a
10\,min temporal resolution. Since foehn winds show a characteristic
wind direction, only observations meeting a specific precondition are
used for classification as described in Section\,\ref{sec:method:mixture}.
Table\,\ref{app:tab:foehnixconfig} shows the wind sectors used for each
target site, as well as the amount of data used for classification.

% latex table generated in R 4.3.1 by xtable 1.8-4 package
% Mon Jun  3 12:25:05 2024
\begin{table}[ht]
\centering
\begin{tabular}{l|p{10cm}}
  \hline
 & Stations, conditional wind direction, data used \\ 
  \hline
Altdorf & Valley station: Altdorf (60$^\circ$--240$^\circ$) \newline Crest station: Gütsch, Andermatt (105$^\circ$--285$^\circ$) \newline 20.0\,\% within sector, 57.1\,\% outside sector, 22.9\,\% removed \\ 
  Innsbruck & Valley station: Universität Innsbruck (130$^\circ$--230$^\circ$) \newline Crest station: Sattelberg (90$^\circ$--270$^\circ$) \newline 5.8\,\% within sector, 49.3\,\% outside sector, 44.9\,\% removed \\ 
  Ellb\"ogen & Valley station: Ellbögen (45$^\circ$--225$^\circ$) \newline Crest station: Sattelberg (90$^\circ$--270$^\circ$) \newline 22.6\,\% within sector, 49.0\,\% outside sector, 28.4\,\% removed \\ 
  Montana & Valley station: Montana (30$^\circ$--100$^\circ$) \newline Crest station: Gütsch, Andermatt (120$^\circ$--180$^\circ$) \newline 23.2\,\% within sector, 52.9\,\% outside sector, 23.9\,\% removed \\ 
  Comprovasco & Valley station: Comprovasco (270$^\circ$--30$^\circ$) \newline Crest station: Gütsch, Andermatt (290$^\circ$--89$^\circ$) \newline 33.5\,\% within sector, 51.2\,\% outside sector, 15.4\,\% removed \\ 
  Lugano & Valley station: Lugano (270$^\circ$--80$^\circ$) \newline Crest station: Gütsch, Andermatt (290$^\circ$--89$^\circ$) \newline 26.9\,\% within sector, 62.1\,\% outside sector, 10.9\,\% removed \\ 
   \hline
\end{tabular}
\caption{Overview of the stations and defined wind sectors used as
precondition for classification. For each target location (left column) the corresponding
valley station and crest station and their required wind sectors
are listed. In addition, the percentage of observations used for classification
(within sector), not used in classification (outside sector) and removed due to missing
data is shown (based on all available data). More station details cf. Table\,\ref{tab:stations}.} 
\label{app:tab:foehnixconfig}
\end{table}
The (conditional) two-component Gaussian mixture model with concomitant variables
\cite{gruen2008,plavcan2014,staufferfoehnix} is defined as follows:
%%%To obtain the probability for foehn occurrence, a two-component Gaussian mixture
%%%model with additional concomitant variables is used (Gr{\"u}n and Leisch 2008).
%%%While the potential temperature difference to the nearby mountain station ($\mathbf{y}$) is used
%%%to model the parameters $\theta$ for the two Gaussian components $f()$ (foehn vs. no foehn),
%%%relative humidity and wind speed are used as concomitant variables $\mathbf{X}$ for modeling
%%%the probability $\pi$ for an observation falling into the second component.
%%%The resulting two-component distribution is thus specified as follows:
$$
h(\mathbf{y}, \mathbf{X}, \mathbf{\theta}, \mathbf{\alpha}) =
    \underbrace{\big(1 - \pi(\mathbf{X}, \mathbf{\alpha})\big) \cdot \mathcal{N}(\mathbf{y}, \mathbf{\theta}_1)}_{\text{first component}} +
    \underbrace{\pi(\mathbf{X}, \mathbf{\alpha}) \cdot \mathcal{N}(\mathbf{y}, \mathbf{\theta}_2)}_{\text{second component}},
$$
where the joint (mixed) density $h()$ is the sum of two Gaussian
components $\mathcal{N}()$ times the probability $\pi()$ that an observation
falls into the second component. The potential temperature $\mathbf{y}$
is used to separate the two components with $\theta_\bullet$ representing
the distribution parameters of $\mathcal{N}()$. The concomitant model
modeling $\pi$ is a logistic regression model employing an intercept as well
as relative humidity and wind speed ($\mathbf{X}$) and can be written as:
$$
\log\big(\frac{\pi}{1 - \pi}\big) = \mathbf{X}^\top \mathbf{\alpha};~~~\pi = \frac{\exp(\mathbf{X}^\top \alpha)}{1 + \exp(\mathbf{X}^\top \alpha)}.
$$
Once the required parameters $\theta$,$\alpha$ are estimated, the final
a-posteriori probability $\hat{p} \in [0,1]$ can be calculated using
$$
\hat{p}(\mathbf{y}, \mathbf{X}, \mathbf{\theta}, \mathbf{\alpha}) = \frac{\pi(\mathbf{X}, \mathbf{\alpha}) \cdot \mathcal{N}(\mathbf{y}, \mathbf{\theta}_2)}{\big(1 - \pi(\mathbf{X}, \mathbf{\alpha})\big) \cdot \mathcal{N}(\mathbf{y}, \mathbf{\theta}_1) + \pi(\mathbf{X}, \mathbf{\alpha}) \cdot \mathcal{N}(\mathbf{y}, \mathbf{\theta}_2)}
$$

The result is the probability
$\hat{p}(\mathbf{y}, \mathbf{X}, \theta, \alpha) = Pr_{\text{obs}}(\text{foehn})$
(Eqn.\,\ref{eqn:foehnevent}) with the same temporal resolution as the
in-situ observations. After the temporal upscaling
(Sec.\,\ref{sec:method:mixture}) this result used to label the data as `foehn'
or `no foehn' event ($\text{Pr}_{\text{1h}}$; Eqn.\,\ref{eqn:method:hourly}).

\section{Supervised learning: Reconstruction}\label{app:sec:method:glm}

The hourly labeled data (Sec.\,\ref{sec:method:mixture}, App.\,\ref{app:sec:method:mixture})
are are combined with the hourly data derived from ERA5
(Sec.\,\ref{sec:data:era5}, App.\,\ref{app:sec:data:era5}), serving
the training data for the supervised learning.

\subsection{lasso: Logistic regression with lasso (L1) regularization}\label{app:model:lasso}

Due to the large number of possible covariates from ERA5 and the fact that many
of these variables are highly correlated, penalization is required.
This study employs lasso (least absolute shrinkage and selection operator) with
L1 regularization, optimizing the following penalized log-likelihood:
%
%logistic regression with L1 regularization is used.
%The regression parameters $\beta$ of the logistic regression model
%\begin{equation}
%    \text{Pr}_{\text{1h}}(\text{foehn}) =
%        \frac{\exp(\text{era5}^\top \beta)}{1 + \exp(\text{era5}^\top \beta)}
%\end{equation}
%are estimated by minimizing the lasso penalized log-likelihood

\begin{equation}
    (\hat{\beta_0},\hat{\beta}) = \underset{(\beta_0,\beta) \in \mathcal{R}^k}{\text{argmin}} \Big[\frac{1}{2N}\sum_{i=1}^{N}(\text{Pr}_{\text{obs},i} - \beta_0 - \text{ERA5}^\top \beta)^2 +
        \lambda ||\beta||_1\Big],
\end{equation}

with $\text{ERA5} \in \mathcal{R}^{N \times k}$, where $N$ is the number of
hourly foehn probabilities (Sec.\,\ref{sec:method:mixture}) and $k$ the
number of available covariates (Sec.\,\ref{sec:data:era5}).

To find the optimal tuning parameter $\hat{\lambda}$, a 30-fold cross-validation
is performed which yields the optimal (regularized) set of regression coefficients
$\hat{\beta_0},\hat{\beta}$ using the \emph{R}\,package \verb|glmnet|
(with \code{s = "lambda.min"}).

\subsection{stabsel: Logistic regression with lasso-based stability selection}\label{app:models:stabsel}

As the lasso penalty shrinks more and more parameters to $0$ with increasing
$\lambda$, this model can also be used for stability selection
\citep{meinhausen2010}. The penalized regression model is estimated $M$ times
for a series of values $\lambda \in [0, \infty]$. The first $K$ parameters
which enter the model ($\ne 0$) will be memorized.

In this study, this is repeated $M = 200$ times, estimating the model on a
random subset with only 50\% of the available data. As the number of `foehn'
events can be rather low (0.6\%--29\% depending on station/time of day)
a stratified random subset (bagging) is
drawn each time containing the same proportion of `foehn' and `no foehn' events
as the full data set. In each of the $M = 200$ iterations, the name of the
first $K = 40$ covariates entering the model is stored. Those covariates which
have been selected more than $60\%$ of the time (at least 121/200 times) are
then used to estimate an additional unregularized logistic regression model.

\begin{enumerate}
    \item Whilst iteration is $\le 200$:
    \begin{enumerate}
        \item Draw stratified subsample of size $N / 2$
        \item Estimate regularized coefficients $(\hat{\beta_0},\hat{\beta})$
            for different $\lambda$s
        \item Extract and keep the name of the first $K = 40$ covariates entering the model
    \end{enumerate}
\item Select the covariates which a selection frequency $>0.6$\label{pseudo:1}
\item Estimate final (unregularized) logistic regression model with the covariates from Step\,\ref{pseudo:1}\label{pseudo:2}
\end{enumerate}

Stability selection allows to automatically identify the most important covariates
among the ones available and to strongly reduce the complexity of the final model
(Step\,\ref{pseudo:2}) without losing much predictive performance.

%%%\begin{leftbar}
%%%    \textbf{TODO(R)}: What we do not mention is what is selected; we would have
%%%    the analysis for that (in groups; direct, temporal, spatio-temporal or in
%%%    physical groups; mass field, \dots).
%%%\end{leftbar}
%%%
%%%\begin{figure}[!ht]
%%%    \includegraphics[width = \textwidth, page = 1]{paper_validation_selection}
%%%    \caption{\textbf{TODO(R)}: I've just left this figure in here as an example of
%%%    the selection analysis.}
%%%    \label{fig:res:selection}
%%%\end{figure}

\subsection{xgboost: Extreme gradient boosting} \label{app:model:xgboost}

Extreme gradient boosting (XGBoost; \citealt{chen2016}) is a supervised machine learning
algorithm using a series of decision trees to best predict the labeled
data (response) using a series of explanatory variables (covariates).
Similar to random forests, XGBoost uses estimates and combines a series of decision
trees (a weak learner). The main difference is that XGBoost uses a gradient decent
method (thus the name) by optimizing the model based on the outcome of the previous
tree by finding additional split points to improve the objective function.

Overly simplified, XGBoost starts with estimating one decision tree and evaluates
its residuals (i.e., the yet unexplained part/misclassified data). These residuals
are then used as input for the next iteration to estimate a new tree to improve
the misclassification from the previous step. 

A series of tuning parameters exist such as the number of iterations to be
performed, the maximum depth of an individual tree, or the learning rate
to update the parameters in each iteration. Choosing these tuning parameters
wrong can lead to overly conservative models (not flexible enough) or models
which overfit the data, both of which can result in a poor (out-of-sample)
predictive performance. One way to avoid these scenarios is to perform
proper cross-validation. We performed a (limited) grid search using
the \emph{R}\,package \verb|xgboost| \citep{chenxgboost} with
the following parameter space:

\begin{itemize}
    \item \verb|eta|, learning rate: $\{0.1, 0.125, 0.15, 0.2\}$
        \label{xgboost:eta}
    \item \verb|max_depth|, maximum depth of each weak learner (tree): $\{5, 10, 20\}$
        \label{xgboost:maxdepth}
    \item \verb|min_child_weight|, minimum sum of instance weight needed in a child: $\{2, 4, 6\}$
        \label{xgboost:minchildweight}
    \item \verb|gamma|, minimum loss reduction required: $\{2, 5, 10\}$
        \label{xgboost:gamma}
\end{itemize}

This yields an overall number of $108$ unique combinations. For each combination,
a 10-fold cross-validation using random stratified subsamples and a maximum of $20$ boosting
iterations is performed. It is then evaluated which parameter set and boosting iteration
resulting in the lowest average out-of-sample error to estimate one final model
using all data. A summary of the selected tuning parameters is shown in 
Table\,\ref{tab:xgboost_paramsummary}.
All models are estimated using the `gbtree' booster and the binomial
logistic objective function.

% latex table generated in R 4.3.1 by xtable 1.8-4 package
% Mon Jun  3 12:25:05 2024
\begin{table}[ht]
\centering
\begin{tabular}{l|r|r|r}
  \hline
 & mean & median & range \\ 
  \hline
subsample & 0.500 & 0.5 & 0.5-0.5 \\ 
  eta & 0.200 & 0.2 & 0.2-0.2 \\ 
  max\_depth & 11.602 & 10 & 5-20 \\ 
  min\_child\_weight & 3.423 & 4 & 2-6 \\ 
  gamma & 2.868 & 2 & 2-10 \\ 
  nround & 19.997 & 20 & 19-20 \\ 
   \hline
\end{tabular}
\caption{Mean, median and range of the selected XGBoost parameters
        via stratified 10-fold cross-validation over all stations and times of the day
        using the `full' variable set.} 
\label{tab:xgboost_paramsummary}
\end{table}

% ===============================================================
% ===============================================================
\section{Long-term reconstruction}\label{app:sec:reconstruction}

\begin{figure}[!h]
    \includegraphics[width = \textwidth]{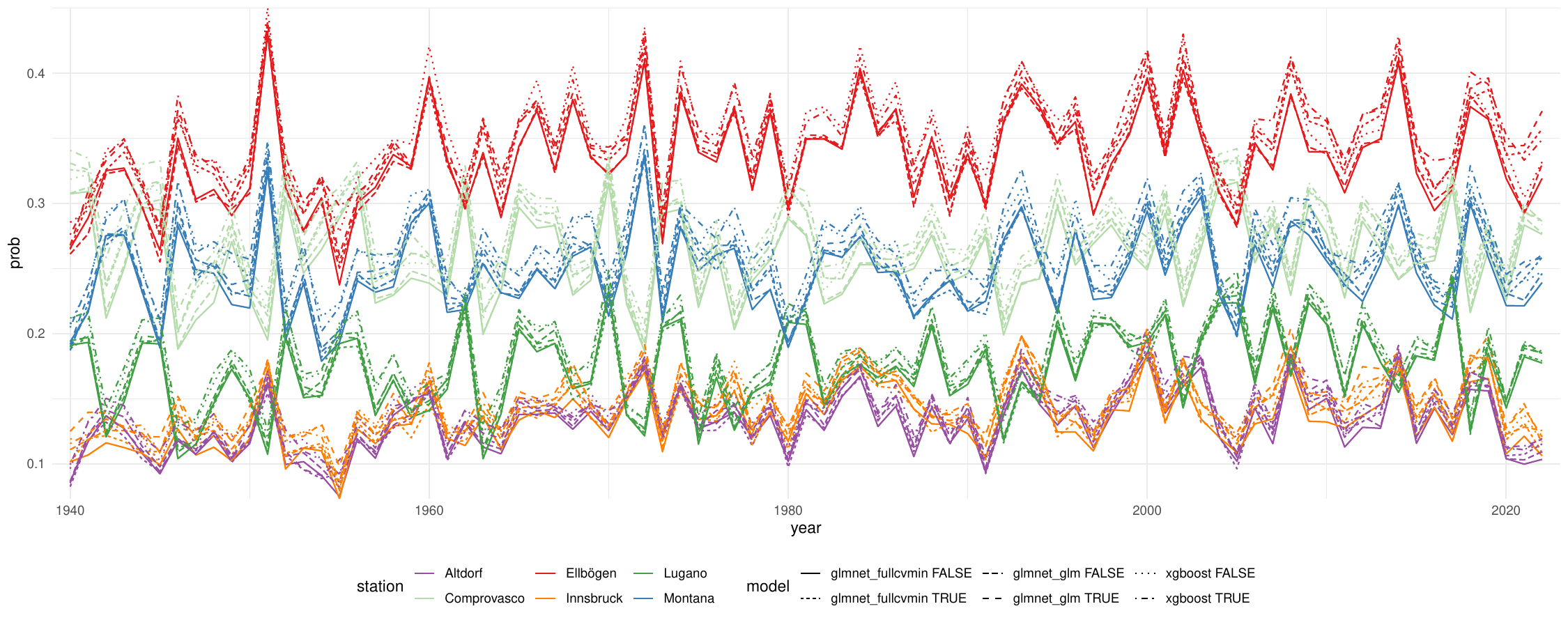}
    \caption{Annual mean of daily maxima as in Figure\,\ref{fig:res:annual}
    for all stations, all three models in both variants.}
    \label{app:fig:res:annual}
\end{figure}

Figure\,\ref{fig:res:annual} (main article, Sec\,\ref{sec:res:annual}) only
shows results of the reconstruction based on the `lasso' model with the `full'
variable set. Figure\,\ref{app:fig:res:annual} shows the same information
for all six stations including the results of all three models (lasso, stabsel, xgboost)
and both variants using either the `direct' variable set or `full' variable set.

The results of these six versions (three models, two variants each) show differences
but agree in large parts, showing similar features across the years.

% ===============================================================
% ===============================================================
\section{Model performance/model comparison}\label{app:sec:modelperformance}

\begin{figure}[!ht]
    \includegraphics[width = \textwidth]{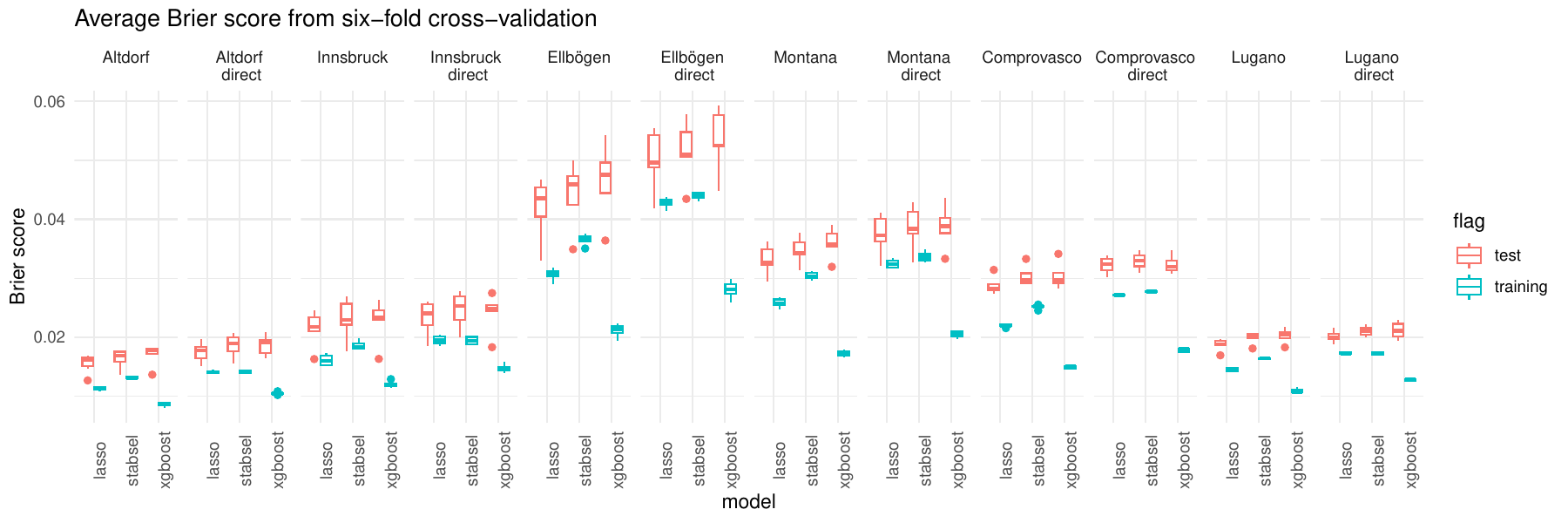}
    \caption{Comparison of average Brier score of all models (lasso, stabsel, xgboost)
    and variants (`direct'/`full') for all six stations.
    Average scores of the six-fold CV are displayed for
    the test (red) and training (green) period.}
    \label{app:fig:res:scores}
\end{figure}

Analogous to Figures\,\ref{fig:res:brier}~and~\ref{fig:res:brierallmodels}
(main article; Sec,\ref{sec:res:fullset}~\&~\ref{sec:res:comparison})
Figure\,\ref{app:fig:res:scores} shows average Brier scores from six-fold CV
for all three models and model variants (using the `direct' or `full' variable set).
As discussed in the main article, the `xgboost' model might be subject to some
overfitting (training) as it always shows lowest scores across all stations and
both model variants.

%%%%Table\,\ref{tab:samplesize} shows the
%%%%number of test and training data for each station and each CV block
%%%%(\textbf{TODO(R):} Left it in here for now but can be dropped I think).

%%%% <<samplesize, results = tex, echo = FALSE>>=
%%%% stopifnot(file.exists(rds <- "paper_tables.rds"))
%%%% res <- readRDS(rds)$samplesize
%%%% res <- res[, -1]
%%%% rownames(res) <- gsub("Ellboegen", "Ellb\\\\\"ogen", rownames(res))
%%%% 
%%%% # Adjusting order to match the plots in the main article
%%%% get_order <- function(x) {
%%%%     # Expected order (regex)
%%%%     e <- c("^Altdorf$", "^Innsbruck", "^Ell", "^Montana$", "^Comprovasco$", "^Lugano$")
%%%%     # Find matches
%%%%     idx <- rep(NA_integer_, length(x))
%%%%     for (i in seq_along(idx)) idx[i] <- grep(e[[i]], x)
%%%%     if (any(is.na(idx))) stop("Problems sorting the table!")
%%%%     return(idx)
%%%% }
%%%% res <- res[get_order(rownames(res)), , drop = FALSE]
%%%% 
%%%% library("xtable")
%%%% res <- xtable(res, caption = "Sample sizes in (test/training, in thousands) used for the cross validation of the logistic regression models.",
%%%%               label = "tab:samplesize")
%%%% print(res,
%%%%       sanitize.rownames.function = function(x) { x })
%%%% @

In addition, event-based metrics are used in the discussion (Sec\,\ref{sec:discussion})
to compare the results of this study to existing literature. 
Table\,\ref{app:tab:scoretable} contains the results for all six stations and both
model variants (`direct'/`full' variable set) for all three supervised methods
(`lasso'/`stabsel'/`xgboost').

% latex table generated in R 4.3.1 by xtable 1.8-4 package
% Mon Jun  3 12:25:05 2024
\begin{table}[ht]
\centering
\begin{tabular}{l|l|r|r|r}
  \hline
Station & Variable set & FNR & FPR & PC \\ 
  \hline
Altdorf & direct & $20.0$/$18.9$/$13.6$ & $0.6$/$0.7$/$0.3$ & $98.4$/$98.3$/$99.0$ \\ 
   & full & $15.7$/$17.1$/$10.7$ & $0.4$/$0.6$/$0.2$ & $98.8$/$98.5$/$99.2$ \\ 
  Innsbruck & direct & $42.9$/$40.6$/$31.7$ & $0.6$/$0.8$/$0.3$ & $97.6$/$97.6$/$98.4$ \\ 
   & full & $37.2$/$36.1$/$24.0$ & $0.5$/$0.8$/$0.2$ & $98.0$/$97.8$/$98.8$ \\ 
  Ellb\"ogen & direct & $19.4$/$19.9$/$12.9$ & $2.9$/$3.0$/$1.6$ & $94.2$/$94.0$/$96.4$ \\ 
   & full & $13.9$/$15.9$/$7.9$ & $1.9$/$2.4$/$1.0$ & $96.0$/$95.3$/$97.8$ \\ 
  Montana & direct & $24.1$/$24.4$/$14.5$ & $1.7$/$1.9$/$0.8$ & $95.7$/$95.5$/$97.6$ \\ 
   & full & $19.4$/$21.7$/$12.1$ & $1.3$/$1.8$/$0.7$ & $96.6$/$95.9$/$98.0$ \\ 
  Comprovasco & direct & $17.4$/$17.6$/$10.3$ & $1.7$/$1.8$/$0.9$ & $96.4$/$96.3$/$97.9$ \\ 
   & full & $13.7$/$15.3$/$8.4$ & $1.3$/$1.6$/$0.7$ & $97.2$/$96.7$/$98.4$ \\ 
  Lugano & direct & $17.2$/$17.3$/$11.8$ & $0.9$/$1.0$/$0.5$ & $97.9$/$97.8$/$98.6$ \\ 
   & full & $15.3$/$16.4$/$9.5$ & $0.7$/$0.9$/$0.4$ & $98.2$/$97.9$/$99.0$ \\ 
   \hline
\end{tabular}
\caption{Miss rate (FNR), false alarm rate (FPR)
and percent correct (PC) in percent based on the entire period (in-sample) for model comparison.
Scores are shown for the `lasso'/`stabsel'/`xgboost' models in this order.
} 
\label{app:tab:scoretable}
\end{table}
\end{appendix}

%% -----------------------------------------------------------------------------
\end{document}